\documentclass[aapm,graphicx]{revtex4-1}

\draft 

\usepackage[mathlines]{lineno}
\modulolinenumbers[5]

\usepackage{amssymb}
\usepackage[pdftex]{graphicx}
\usepackage{multirow}
\usepackage{color,soul}

\begin{document}

\title{Dosimetric performance of a multi-point plastic scintillator dosimeter as a tool for real-time source tracking in high dose rate brachytherapy} 

\author{Haydee M. Linares Rosales$^{1,2}$}
\author{Louis Archambault$^{1,2}$}
\author{Sam Beddar$^{3, 4}$}
\author{Luc Beaulieu$^{1,2}$}

\affiliation{$^{1}$D\'epartement de physique, de g\'enie physique et d'optique et Centre de recherche sur le cancer, Universit\'e Laval, Qu\'ebec, Canada.}
\affiliation{$^{2}$D\'epartement de radio-oncologie et Axe Oncologie du CRCHU de Qu\'ebec, CHU de Qu\'ebec - Universit\'e Laval, QC, Canada.}
\affiliation{$^{3}$Department of Radiation Physics, The University of Texas MD Anderson Cancer Center, Houston,TX, United States.}
\affiliation{$^{4}$The University of Texas MD Anderson UTHealth Graduate School of Biomedical Sciences, Houston, TX, United States.}

\email[Corresponding author: Haydee M. Linares Rosales, ]{haydee8906@gmail.com}
\date{\today}

\begin{abstract}
\scriptsize{\textbf{Purpose:} This study aims to present the performance of a multi-point plastic scintillation detector (mPSD) as a tool for real-time dose measurements (covering three orders of magnitude in dose rate), source-position triangulation, and dwell time assessment in high dose rate (HDR) brachytherapy.\\

\textbf{Methods}: A previously characterized and optimized three-point sensor system was used for in vivo HDR brachytherapy measurements. The detector was composed of three scintillators: BCF-60, BCF-12, and BCF-10. Scintillation light was transmitted through a single 1-mm-diameter clear optical fibre and read by a compact assembly of photomultiplier tubes (PMTs). Each component was numerically optimized to allow for signal deconvolution using a multispectral approach, taking care of the Cerenkov stem effect as well as extracting the dose from each scintillator. The PMTs were read simultaneously using a data acquisition board at a rate of 100 KHz and controlled with in-house software based on Python. An $^{192}Ir$ source (Flexitron, Elekta-Brachy) was remotely controlled and sent to various positions in a home-made PMMA phantom, ensuring 0.1 mm positional accuracy. Dose measurements covering a range of 0.5 to 10 cm from the source were carried out according to TG-43 U1 recommendations. Water measurements were performed in order to: (1) characterize the system’s response in terms of angular dependence; (2) obtain the relative contribution of positioning and measurement uncertainties to the total system uncertainty; (3) assess the system’s temporal resolution; and (4) track the source position in real time. The triangulation principle was applied to report the source position in three-dimensional space.\\

\textbf{Results:} As expected, the positioning uncertainty dominated close to the source, whereas the measurement uncertainty dominated at larger distances. A maximum measurement uncertainty of 17 \% was observed for the BCF-60 scintillator at 10 cm from the source. Based on the uncertainty chain, the best compromises between positioning and measurement uncertainties were reached at 17.2 mm, 17.4 mm, and 17.5 mm for the BCF-10, BCF-12, and BCF-60 scintillators, respectively, which also corresponded to the recommended optimal distances to the source for calibration purposes. The detector further exhibited no angular dependence. All dose values were found to be within 2\% of the dose value at 90$^{\circ}$. In the experiments performed for source-position determination, the system provided an average location with a standard deviation under 1.7 mm. The maximum observed differences between measured and expected values were 1.82 mm and 1.8 mm in the x- and z-directions, respectively. Deviations between the mPSD measurements and expected TG-43 values were below 5\% in all the explored measurement conditions. With regard to dwell time measurement accuracy, the maximum deviation observed at all distances was 0.56 $\pm$ 0.25 s, with a weighted average of the three scintillators of 0.07 $\pm$ 0.05 s at all distances covered in this study. \\

\textbf{Conclusions:} Real-time HDR brachytherapy measurements were performed with an optimized mPSD system. The performance of the system demonstrated that it could be used for simultaneous, in vivo, real-time reporting of dose, dwell time, and source position during HDR brachytherapy.}

\end{abstract}


\maketitle

\section{Introduction} 

High dose rate (HDR) brachytherapy is a radiation therapy procedure in which the radioactive sources are placed a short distance from the target. This modality is characterized by a high dose gradient near the source (20\%/mm or more for the first centimeter), a feature that affords a high level of protection to surrounding healthy tissues. Owing to these high dose gradients, small uncertainties can result in significant dose variations. Thus, if small errors take place during the treatment and are not immediately detected, harmful consequences and secondary radiation effects may occur. If detected at all, these errors are typically only identified after treatment because of the limited availability of commercial real-time treatment-monitoring systems. Afterloader safety systems can identify dose delivery errors that originate from mechanical obstruction of the source and improper guide-tube connections. However, incorrectly specified source strengths or erroneously connected source-transfer guide tubes can go unnoticed \cite{Kertzscher-2011, ICRP-97-Brachy}.

Routine in vivo dosimetry can be a powerful tool to determine whether deviations from the treatment plan occur during treatment delivery. In vivo dosimetry provides direct information about the level of agreement between planned and measured doses in or near the tumor region. However, it requires a radiation detection system capable of measuring the cumulative dose or dose rate with good sensitivity, precision, and accuracy. Different types of detectors have been studied for in vivo dosimetry applications in brachytherapy  \cite{DAS-TLD-invivo-2007, Toye-TLD-invivo-2009, Anagnostopoulos-TLD-invivo-2003}. A review by Tanderup \textit{et al.} \cite{Tanderup-invivo-Brachy-2013}  highlighted the main aspects of various detectors that could be used as in vivo dosimeters in brachytherapy. One such detector, the plastic scintillation detector (PSD), has several advantages that have been recently highlighted in the literature, a key one being their real-time response \cite{Therriault-Temp-method-2015, Boivin-2016, Lambert-Cerenkov-2008, Beddar-Cerenkov-1992, Archambault-2006, Wootton-Temperature-2013, Guillot-toward-2010, Beddar-water-equivalent-1992-1, Beddar-water-equivalent-1992-2, Beaulieu-Scint-Status-2013, Beaulieu-Review-2016, Linares-2019}. Although PSDs are affected by the stem effect and temperature variations \cite{Wootton-Temperature-2013, Beddar-temp-2012}, several investigations have developed methods to correct both of these dependencies in the detector response  \cite{Beddar-water-equivalent-1992-1, Beddar-water-equivalent-1992-2, Boer-optical-1993, Fontbonne-2002, Clift-temporal-2002, Lambert-Cerenkov-2008, Archambault-MathForm-2012,Therriault-Temp-method-2015}. 

Most of the studies characterizing PSD response were conducted using an optical fibre connected to a single point of measurement as a sensitive volume. However, studies have also demonstrated the feasibility of using multiple scintillation detectors (mPSDs) attached to a single optical chain \cite{Archambault-MathForm-2012, Therriault-mPSD-2012}. A study done by Linares Rosales et \textit{al.} \cite{Linares-2019} characterized the response of an mPSD system for application to HDR brachytherapy; the authors demonstrated that with proper optimization of the signal collection chain, this mPSD system is accurate within clinically relevant distances from the source. Additionally, previous work explored the source-tracking capacity of different detectors in HDR brachytherapy \cite{Therriault-mPSD-Brachy-2013,Johansen-2018,Smith-2016,Guiral-2016,Nakano-2005,Fonseca-2017}. Some studies used an array of dosimeters placed on the patient’s skin, and others a flat-panel detector. In a study of source-position tracking with a single-point detector in HDR brachytherapy, Johansen \textit{et al.} \cite{Johansen-2018} used the dose values from the treatment planning system to develop a method to determine average source shifts within catheters through a Gaussian fit. Besides the aforementioned, brachytherapy clinics do not verify their treatments in real time.  The available real time systems present small signal-to-noise ratios, limited time resolution, large measurements uncertainties and can detect only errors in the order of 20\% or more \cite{Kertzscher-2016-Ruby}. The current study presents the dosimetric performance of a previously optimized and characterized mPSD system in the context of in vivo dosimetry for HDR brachytherapy. Through in-water dose measurements, we: (1) evaluated the angular response of the dosimeter; (2) determined the relative contribution of positioning and measurement uncertainties to the total uncertainty chain; (3) assessed the capacity of the system to measure individual dwell times; and (4) tracked the source position in real time. 

\section{Materials and Methods}

\subsection{A mPSD system components}

\label{MM_mPSD_dosim_sys}

\begin{figure}
\centering
\begin{tabular}{c}
\includegraphics[trim = 0mm 0mm 0mm 0mm, clip, scale=0.75]{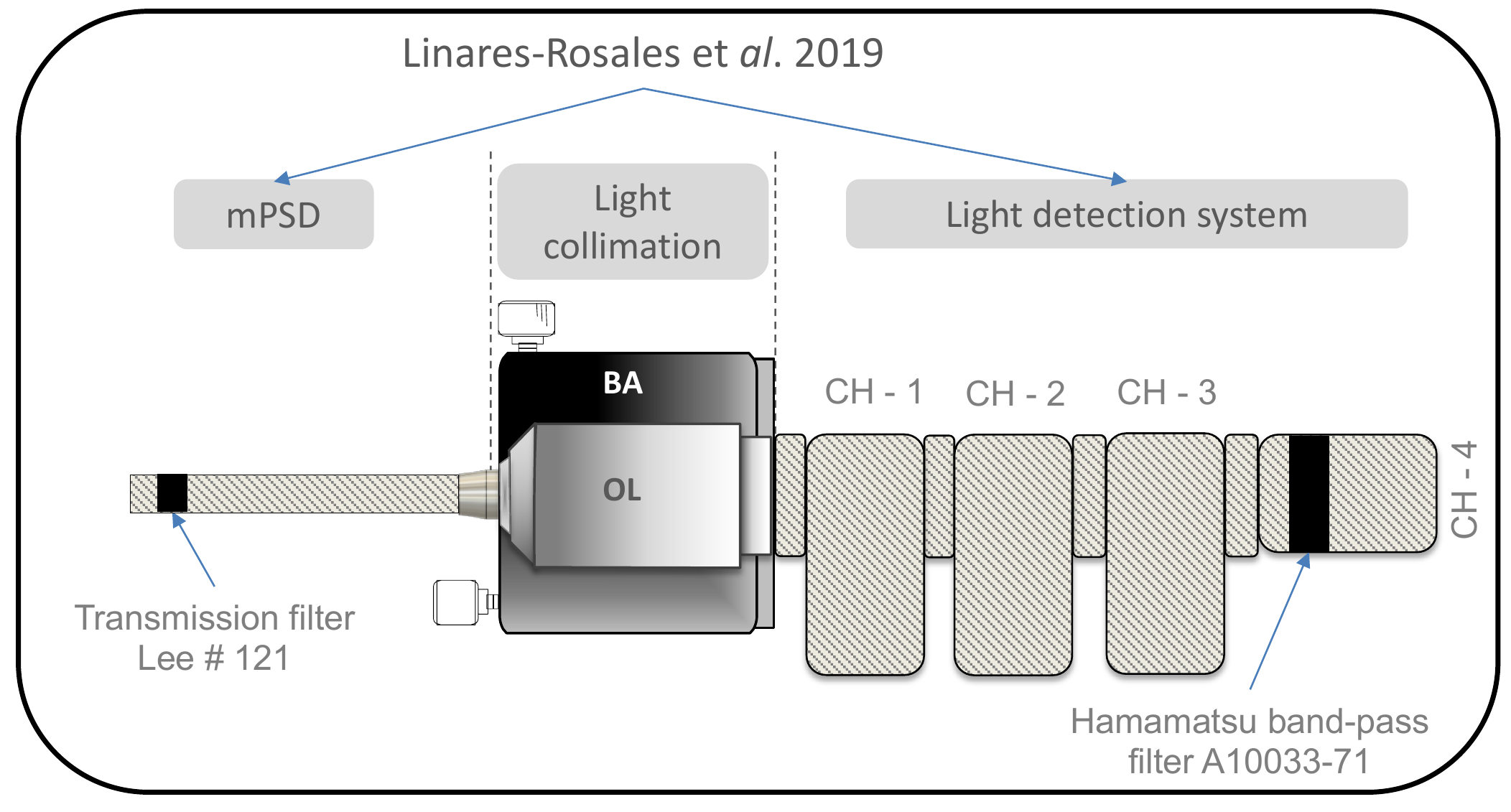}
\end{tabular}
\caption{\label{mPSD_luminescence_system} Schematic of the dosimetry system used for HDR brachytherapy dose measurements. The cross-hatched components represent similarities with the system reported by Linares Rosales et \textit{al.} \cite{Linares-2019}. A green transmission filter was placed after the BCF-60 PSD to avoid cross-excitation from the BCF-10 and BCF-12 PSDs. CH indicates measurement channels. The filter used in CH-4 was the A10033-71 from Hamamatsu. BA refers to beam aligner block A10760 from Hamamatsu. An Olympus infinity-corrected objective lens (OL), model RMS40X from Thorlabs, was coupled to the BA block.}
\end{figure}

The scintillation light is generated in a three-point PSD and detected through photomultiplier tubes (PMTs) coupled to a set of dichroic mirrors and filters, resulting in a combination that allows for the deconvolution of scintillation light into different spectral bands. Figure \ref{mPSD_luminescence_system} shows a schematic of the dosimetry system used in this study, which is similar to the system reported by Linares Rosales et \textit{al.} \cite{Linares-2019}. 
The cross-hatched components in figure \ref{mPSD_luminescence_system} represent the components that were also used in that system.  Each assembly, composed of a dichroic mirror, filter, and PMT, is referred as a channel (CH). According to the hyperspectral filtering technique proposed by Archambault et al. \cite{Archambault-MathForm-2012},the number of channels to be used depends on the number of scintillator points N composing the mPSD, and equals $N + 1$. The additional channel is used to take into account the stem effect, which must be removed from the measured signal \cite{Therriault-2011}.

A few key changes were made to the system to improve its performance and obtain higher overall light-collection efficiency. First, a filter with a transmission spectrum in the range of 475 to 600 nm was added to the mPSD after the BCF-60 scintillator because a measured residual angular effect came from cross-excitation of the BCF-10 and BCF-12 scintillators. This effect is characterized in Section \ref{Result_Ang_Dependence}. The chosen filter was the Lee filter \#121 from PNTA (Seattle, WA, USA). The coupling technique used for detector construction was previously described by  Ayotte et \textit{al.}  \cite{Ayotte-Surface-2006}. 
Second, a beam aligner block (BA; module A10760 from Hamamatsu, Bridgewater, NJ, USA) \cite{PMT-Hamamatsu}) was included at the entrance of the light-collection system, coupled to an Olympus infinity-corrected objective lens (OL; RMS40X from Thorlabs, Newton, NJ, USA). Note that the filter in CH-4 is also different from that initially recommended by Linares Rosales et \textit{al.} \cite{Linares-2019}. Section \ref{MM_Improv_Light_Coll} describes the experiments performed to evaluate the impact of these new components on the light-collection efficiency.

The detector was made light-tight to avoid environmental light contribution and physical damage. The mPSD's 1-mm inner diameter allowed its insertion into a 30-cm needle set from Best Medical International (Springfield, VA, USA), which was used during measurements. Furthermore, all the components were enclosed in a custom-made black box to exclude external light.

A data acquisition board (DAQ NI USB-6289 M Series Multifunction I/O Device from National Instruments, Austin, TX, USA) \cite{DAQ-6289} read the signal produced in each channel at a rate of 100 kHz and sent it to a computer (Apple MacBook Pro, 2.9 GHz Intel Core i5). The light-detection system was controlled independently from the irradiation unit with in-house software based on Python.

\subsection{Performance of light collection apparatus}
\label{MM_Improv_Light_Coll}

\begin{figure}
\centering
\begin{tabular}{c}
\includegraphics[trim = 0mm 0mm 0mm 0mm, clip, scale=0.55]{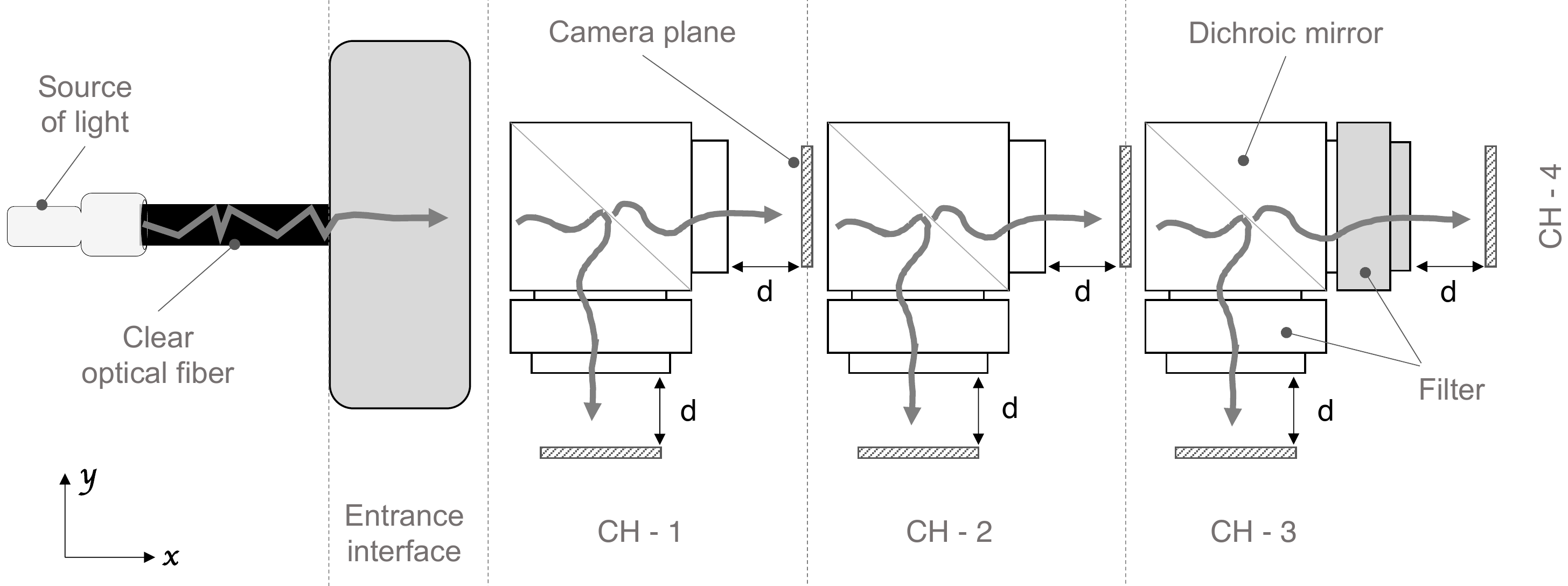}
\end{tabular}
\caption{\label{Improved_syst_setup} Schematic of the experimental set-up used to evaluate the light-collection efficiency at each step. The shaded regions are the major component changes introduced in this dosimetry system. The entrance interface is the region where the connection between the fiber and the first channel takes place. CH, channel.}
\end{figure}

Figure \ref{Improved_syst_setup} shows a schematic of the experimental set-up used to evaluate the effect of using the BA block and the A10033-71 filter. The shaded components in  figure \ref{Improved_syst_setup} highlight the changes introduced in the dosimetry system used in this study from that used by Linares Rosales et \textit{al.} \cite{Linares-2019}. A white light source (model HL-2000 from Ocean Optics, Dunedin, USA) was fixed at one end of a clear optical fiber (Eska GH-4001 from Mitsubishi Rayon Co., Ltd., Tokyo, Japan), while the other end was connected to the system entrance interface. The connection between the fiber and the first channel was named the “entrance interface” to highlight that two types of components were used in that space: (a) a subminiature version SMA adaptor like that used by Linares Rosales \cite{Linares-2019} and (b) the BA block. As shown in Figure \ref{Improved_syst_setup} , the light passes through the entrance interface and strikes a dichroic mirror. Depending on the properties of the dichroic mirror, some of the incoming light is transmitted in the x-direction, while the reflected light goes in the y-direction, passing through a bandpass filter. The transmitted light then reaches a second and a third dichroic mirror, each with different reflection and transmission properties. The amount of light being transmitted or reflected was quantified in every interface. Thus, we were able to characterize both the light-collection efficiency at each plane and the divergence of the light beam. For this analysis, we replaced the PMTs from the original system with a charge-coupled device camera (Alta U2000, Apogee, Roseville, CA, USA). Each channel’s output was set at a fixed distance, d, of 80 cm from the camera. Ten images were acquired in two planes for each CH module, as shown in Figure \ref{Improved_syst_setup}, and the background signal was subtracted.

To characterize the light divergence, we obtained the full width at half maximum (FWHM) on each picture profile. The light-collection efficiency was evaluated through the collected light intensity in the profile’s plateau. The system developed by Linares Rosales et al. \cite{Linares-2019} without any modifications was the reference system for the quantification of the signal-collection efficiency. Two tests were done to perform this quantification. In Test \#1, we solely evaluated the effect of using the BA block at the entrance interface, without any further modification to the Linares Rosales system. Test \#2 evaluated the impact of the A10033-71 filter on the light collected in CH-4. The BA block was used in the entrance interface.

\subsection{HDR brachytherapy irradiation unit}

Dose measurements were carried out using a Flexitron HDR afterloader from Elekta (Elekta Brachy, Veenendaal, The Netherlands). The cylindrical $^{192}Ir$ source pellet was 0.6 mm in diameter and 3.5 mm in length and was housed inside a stainless steel capsule of 0.86 mm diameter and 4.6 mm length. The source air kerma strength (Sk) was 43810 U. The HDR brachytherapy unit was remotely controlled and able to move the source to the desired position in a water tank by means of a 30-cm needle set from Best Medical International. The mPSD was inserted into an additional catheter for use in real-time dose verification.

\subsection{System calibration, dose measurements and Cerenkov removal}

Dose values were recorded in real time by the mPSD under full TG-43 U1 conditions \cite{TG-43-Update}. All measurements were repeated at least five times, and the set-up was completely disassembled and reassembled between measurements. The mathematical formalism proposed by Archambault et al. \cite{Archambault-MathForm-2012} was used to remove the stem signal.  The calibration matrix and dose values were calculated according to the formulation published by Linares-Rosales et \textit{al.} \cite{Linares-2019} for a 3-points mPSD configuration. Calculations were done with a coordinate system, where the radial direction to the source was represented as $x$ and the longitudinal direction as $z$. Calibrations and measurements were carried out under the same experimental conditions. During the calibration process, measurements were performed with the $^{192}Ir$ source dwelling inside the catheters with a 1-mm step, and the detector positioned at a known $x$-distance from the measurement catheter.  Thus, the source dwell position, where the maximal signal was produced, was related to each sensor $z = 0$ coordinate. Therefore, the relationship between the produced signal and the TG-43 dose was derived, being the calibration matrix independent of detector positioning errors. The absorbed dose deviations for the mPSD were evaluated using the dose predicted by the TG-43 U1 formalism \cite{TG-43-Update} as the reference. Dose values provided by the scintillators were integrated over the scintillator volume to account for their finite size. 

\subsection{Relative contribution of the positioning and the measurement uncertainties}

\label{MM_U_contribution}

The proper selection of calibration conditions is important for measurements of detector response and performance: agreement with the TG-43 U1 expected dose, angular dependence, and signal-to-noise ratio. The selection of the calibration distance was a compromise between measurement uncertainties and positioning errors. Andersen \textit{et al.} in 2009 \cite{Andersen-time-resolved-2009} showed that positioning uncertainty dominates in measurements made close to the source, whereas measurement uncertainty dominates at large distances. We performed an uncertainty analysis similar to the one described by Andersen et \textit{al.} \cite{Andersen-time-resolved-2009} to select the most effective  calibration distance for the mPSD.

Dose as a function of distance to the source as predicted by TG-43 U1 for each scintillator constituted the reference dose. The uncertainty associated with the reference dose was obtained by calculating the dose gradient per millimeter. This uncertainty is represented as $U_{TPS}$. Dose measurements were associated with a standard uncertainty called $U_{M}$. We estimated $U_{M}$ by taking 10 different measurements, each one with a dwell time of 30 s per source position. Thus, at each explored source dwell position, $U_{M}$ was determined using a sample of 300 measurements. Knowing the relative contribution of the positioning uncertainty $U_{TPS}$ and the measurement uncertainty $U_{M}$ , we were able to estimate the combined uncertainty $U_{C}$ associated with each scintillator as a function of the source-to-detector distance. The point where the combined overall uncertainty was the smallest is called in this paper the ''sweet spot'' and was the distance chosen for each independent scintillator's calibration.

\subsection{Angular dependence}

We next explored the variation in the mPSD's response as a function of variation in the angle to the HDR brachytherapy source. Because the shape of the scintillators used was cylindrical, no axial angle dependence was expected, but longitudinal angle dependence was possible.

\begin{figure}
\centering
\begin{tabular}{c}
\includegraphics[trim = 5mm 1mm 1mm 1mm, clip, scale=0.45]{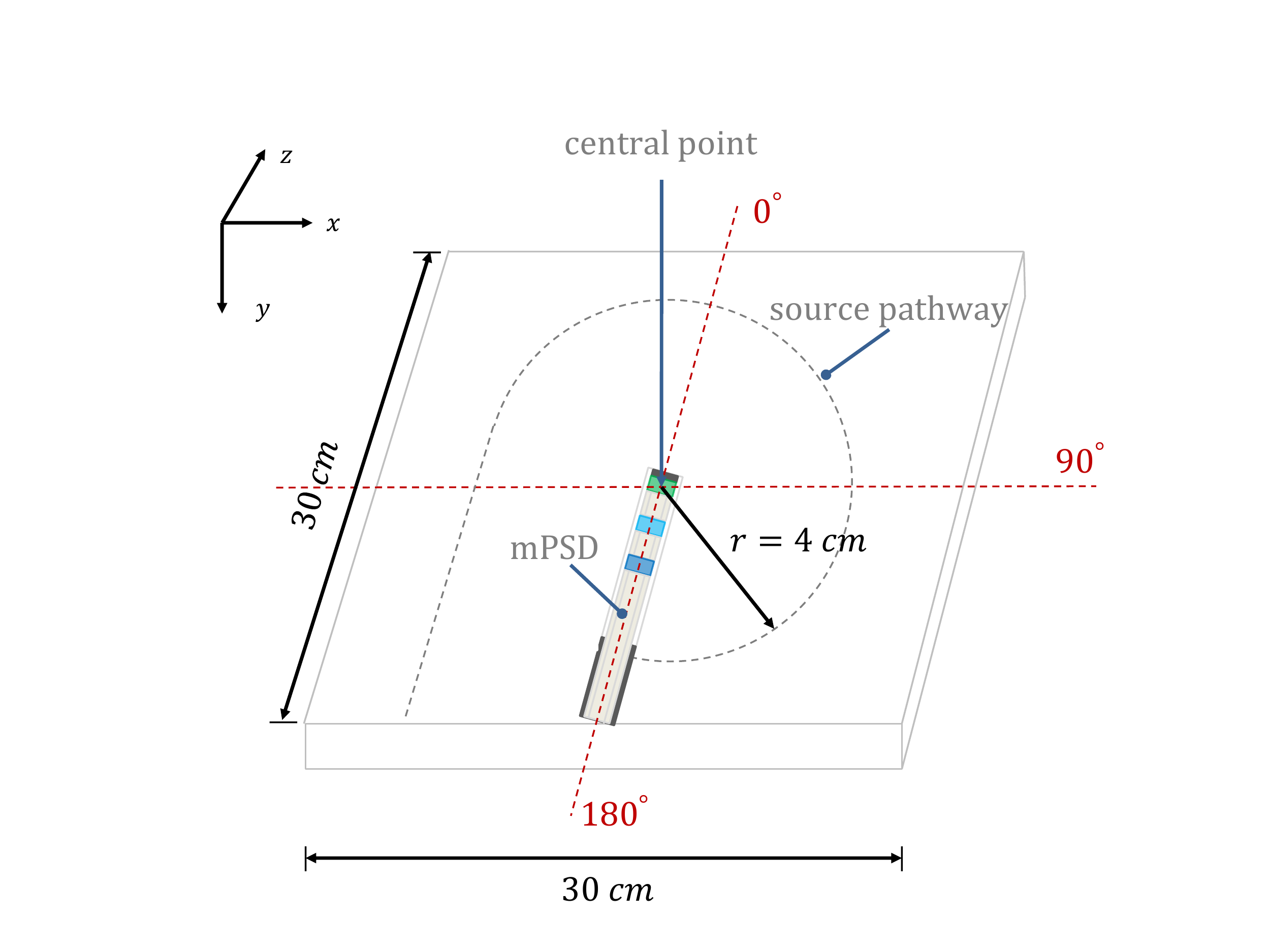}
\end{tabular}
\caption{\label{mPSD_angle} Schematic representation of the template used for mPSD angular dependence analysis.}
\end{figure}

Precise detector positioning is key when evaluating a detector’s angular response. In this work, the scintillation detector was precisely positioned by using a home-made template, as shown in Figure \ref{mPSD_angle}. It consisted of a solid-water slab of 30 $\times$ 30 $\times$ 1 $cm^{3}$. The source catheter lay in a groove in the template slab. The groove radius was 4 $cm$, allowing for a 270$^{\circ}$ source-rotation angle around the mPSD. The source was sent to each specific position using a flexible catheter (LumenCare Azure 5F (Nucletron, Veenendaal, The Netherlands). A source dwell time of 20 s was planned at each position. The slab containing the angular variation template was submerged in a 40 $\times$ 40 $\times$ 40 $cm^{3}$ water tank.  To confirm the source position, we performed initial irradiations with an EBT3 film. The angular dependence study was conducted by placing the sensor at the center of the template (see figure \ref{mPSD_angle}). Repeated irradiations were performed keeping all other variables fixed.  To evaluate the effect of adding a green filter to the BCF-60 scintillator (see Figure \ref{mPSD_angular_dep}) , two mPSDs were used: (1) an mPSD assembly with a green filter coupled to BCF-60, as shown in Figure  \ref{mPSD_luminescence_system}; and (2) an mPSD assembly with no filter but with the exact same physical characteristics. Each measurement was acquired five times, and the set-up was completely unmounted between measurements. In addition, the same procedure was repeated on three different days.

\subsection{$^{192}Ir$ source tracking}

We then evaluated the mPSD system’s ability to report the position of the source in three-dimensional space. This study was done with a precalibrated system under full TG-43 U1 conditions \cite{TG-43-Update}. 

Since the mPSD was held straight inside a catheter and the distance between the scintillators was known, it was possible to apply the triangulation principle to determine the source position relative to the mPSD. The cylindrical geometry of the sensitive volumes in the mPSD allowed for source-position reporting with degeneration in the detector's radial direction. The direct relationship between the dose and distance to the $^{192}Ir$ source was used to build each scintillator response function used for source position triangulation. The voltages produced at each channel with the source at various positions were recorded in real time and translated into a dose value. Then, these measured dose values were introduced as input information in the scintillators’ dose response function and interpolated to determine the distance to the source. As the mPSD was composed of three scintillators, three combinations of source coordinates ($x_{i}, z_{i}$) were determined for each dwell position. The overall source position in space ($\overline{x}, \overline{z}$) was determined through a weighted average calculation, as shown in equations \ref{x_calc} and \ref{z_calc}. In equations \ref{x_calc} and \ref{z_calc}  $\sigma_{i}$ refers to the combined standard deviation associated with the source-scintillator distances.

\begin{equation}
\label{x_calc}
 \overline{x} = \frac{\sum\limits_{i} x_{i}/ \sigma_{i}^2}{\sum\limits_{i} 1/\sigma_{i}^2}
\end{equation}
\begin{equation}
\label{z_calc}
 \overline{z} = \frac{\sum\limits_{i} z_{i}/ \sigma_{i}^2}{\sum\limits_{i} 1/ \sigma_{i}^2}
\end{equation}

\begin{figure}
\centering
\begin{tabular}{c}
\includegraphics[trim = 1mm 1mm 1mm 1mm, clip, scale=0.47]{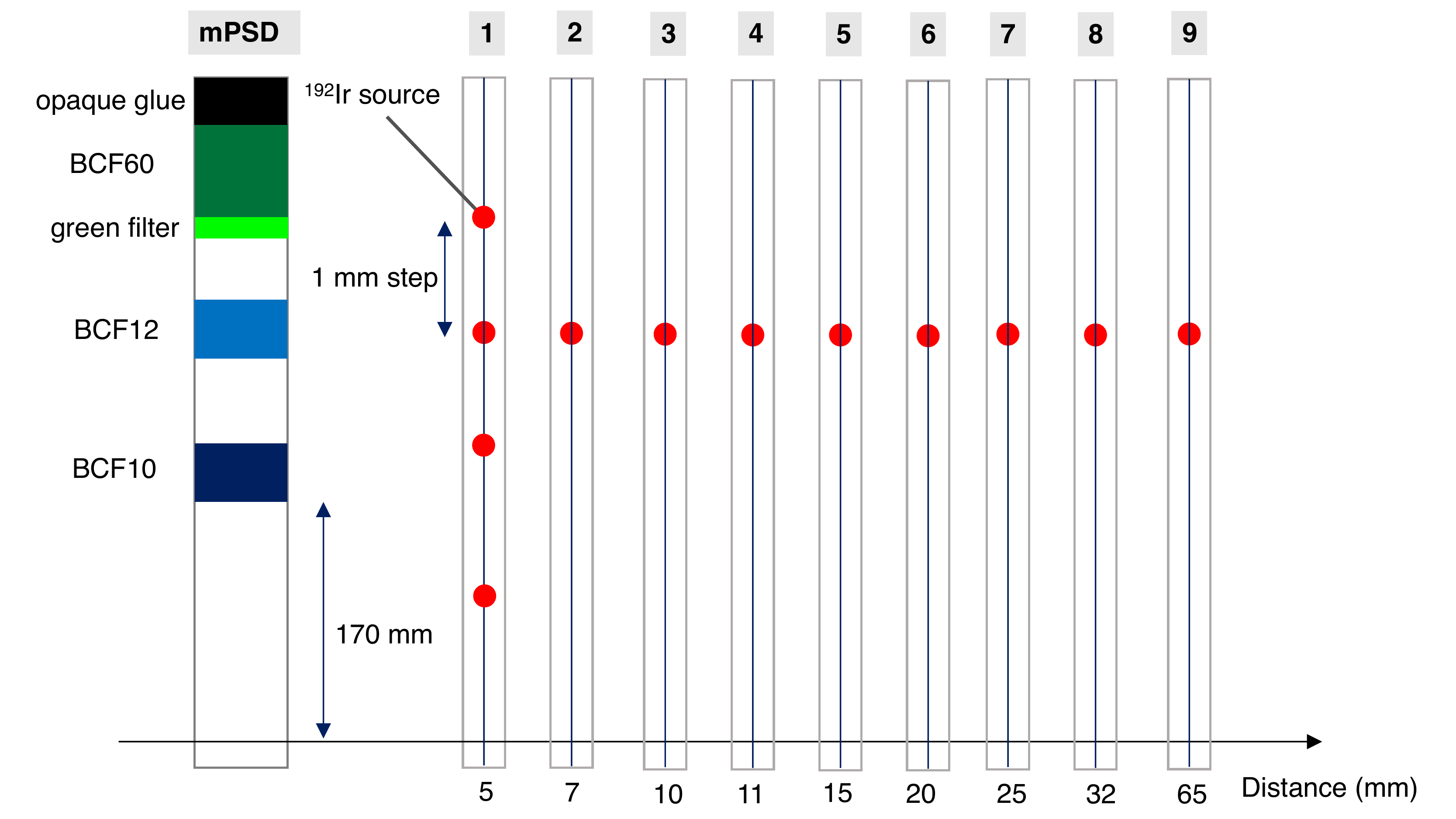}
\end{tabular}
\caption{\label{Scheme_tracking} Schematic of the nine catheters and source positions used for source positioning tracking with the mPSD.}
\end{figure}

Real-time measurements were acquired while the source dwelled at different distances from the mPSD, and an off-line analysis was performed. Measurements were performed with the source and detector isotropically covered by at least 20 cm of water to ensure a full scatter condition, as required by the TG-43 U1 formalism \cite{TG-43-Update}. The catheters were inserted in a custom-made poly(methyl methacrylate) phantom composed of two catheter insertion templates of 12 x 12 $cm^2$, separated by 20 cm \cite{Therriault-mPSD-Brachy-2013,Linares-2019}. This phantom was placed inside a 40 x 40 x 40 $cm^{3}$ water tank to mimic TG43 U1 conditions for a high energy source (i.e. 20 cm of water past the last measurement position \cite{Perez-2012-HEBD}), allowing for source-to-detector parallel displacement with 0.1-mm positioning accuracy. Figure \ref{Scheme_tracking} is a schematic representation of the plans created to track the source position. Nine needles were used to send the source to the desired position. The numbers at the top of Figure \ref{Scheme_tracking} are the catheter numbers, while those at the bottom indicate the distance from the source to the mPSD.

Two irradiation plans were created to test the ability of the system to track the source position. In Plan 1, the source dwelled only inside Catheter 1, with a 1-mm step between each dwell position. In total, the plan had 101 dwell positions with a dwell time of 10 s each. In Plan 2, the HDR source dwelled inside Catheters 1 through 9, but only once per catheter. As in Plan 1, a dwell time of 10 s per source position was planned. Both Plans 1 and 2 were delivered seven times each.

\subsection{Planned vs. mPSD's measured dwell time}

We further evaluated the ability of the real-time mPSD measurements to extract dwell times under various irradiation conditions. Seven irradiation plans were created, and all the parameters of the plan were fixed except the dwell time. In all cases, the source dwelled inside Catheter 1 (Figure \ref{Scheme_tracking} with a 1-mm step between each position. The dwell times used were 1, 2, 3, 4, 5, 10, and 20 s.

For the signal pulse produced at a planned dwell position ($dp$), dwell times were extracted from measurements using the following parameters \cite{Linares-2019}: (a) mean signal ($\mu_{s}$); (b) mean background signal ($\mu_{b}$); (c) signal standard deviation ($\sigma_{s}$); and (d) background standard deviation ($\sigma_{b}$). An active dwell position was considered when $\mu_{s} \pm \sigma_{s} > \mu_{b} \pm \sigma_{b}$. To distinguish the signal from one dwell position $dp_{N}$ from that of the subsequent one $dp_{N+1}$, we considered as a discriminator the relationship $(\mu_{s, dp_{N}} \pm \sigma_{s, dp_{N}}) \ne (\mu_{s, dp_{N+1}} \pm \sigma_{s, dp_{N+1}})$. Once the dwell position $dp_{N}$ from the whole collected signal was isolated, the measured elapsed time was quantified.  Discrepancies in dwell time measurements were evaluated using the planned dwell times as references.

\section{Results and Discussions}

\subsection{Improved light collection efficiency}
\label{Result_Improv_Light_Coll}

Table \ref{Beam_aligner_results} summarizes the results obtained from our experiment investigating the impact of the BA block on the efficiency of signal collection. The first column in Table \ref{Beam_aligner_results} shows the location where the images were acquired according to the schematic shown in Figure \ref{Improved_syst_setup}. Columns 2 and 3 show the mean signal intensity ($\mu_{s}$) obtained in the profile’s plateau region as well as its associated standard deviation $\sigma$. The $\mu_{s}$ values were normalized to the image $\mu_{s}$ value obtained at the fiber's output. Column 4 shows the gain factor at each interface. Columns 5 and 6 show the profile's FWHM obtained for the reference system and Test \# 1, respectively. Because of the geometry and light-cone divergence, signal losses were observed in all the channels. Nonetheless, these results demonstrate that the additional optical block helps to collimate the light transmitted through the mPSD’s optical fiber and consequently reduce the signal collection losses by a factor of almost 2. Important gains were observed in CH-2’s x and y-directions and CH-3, reaching values of 3.26, 4.19, and 4.60, respectively. According to the analysis of the profile’s FWHM, the mean FWHM value for the reference system was 4.43 $\pm$ 0.24 mm, while in Test \#1 it was 4.06 $\pm$ 0.08 mm. 

\begin{table}
\caption{\label{Beam_aligner_results} Results of the analysis of the beam aligner (BA) block’s effect on the dosimetry system. $\mu_{s}$ refers to the mean signal obtained in the profile’s plateau, and $\sigma$ its associated standard deviation. $\mu_{s}$ values are normalized to the image  $\mu_{s}$ obtained at the entrance interface’s output.}
\centering
\vspace{0.2cm}
	\begin{tabular}{cccccc}
	\hline
	\hline
	 & \multicolumn{3}{c}{\textbf{Normalized signal intensity}} & \multicolumn{2}{c}{ \textbf{FWHM}} \\
	 & \multicolumn{3}{c}{\textbf{($\mu_{s}\pm \sigma$)}} & \multicolumn{2}{c}{ \textbf{(mm)}} \\
	 \textbf{Location}	&	\textbf{Ref. System}	&	\textbf{Test \#1}	&\textbf{Gain} &	\textbf{Ref. System}	&	\textbf{Test \#1}	\\
	\hline
	Entrance interface	&	0.510 $\pm$ 0.11	    &	0.88 $\pm$ 0.08	    & 1.73 &	4.81	&	4	\\
    CH - 1 x	&	0.421  $\pm$ 0.09	&	0.760 $\pm$ 0.10        & 1.79 &	4.81	&	4	\\
    CH - 1 y	&	0.118 $\pm$ 0.19	&	0.180 $\pm$ 0.01        & 1.53 &	4.46	&	4.1	\\
    CH - 2 x	&	0.092 $\pm$0.15	    &	0.300 $\pm$ 0.12        & 3.26 &	4.65	&	4.1	\\
    CH - 2 y	&	0.020 $\pm$ 0.18	&	0.084 $\pm$ 0.14        & 4.19 &	4.27	&	4	\\
    CH - 3	    &	0.002 $\pm$ 0.12	&	0.010 $\pm$ 0.12	    & 4.60 &	4.21	&	4	\\
    CH - 4	    &	0.040 $\pm$ 0.14	&	0.068 $\pm$ 0.13        & 1.70 &	4.23	&	4.21	\\
	\hline
	\multicolumn{6}{c}{\small{\textit{FWHM: Full width at half-maximum, CH: Channel}}}
	\end{tabular}
\end{table}

The results from Test \# 2 are not shown in Table \ref{Beam_aligner_results} because changing the filter in CH-4 only influenced the light collected in that channel. The mean signal obtained was 0.63 $\pm$ 0.12, in contrast to the mean signal of 0.35 $\pm$ 0.13 obtained in Test \#1. A10033-63 and A10033-71 are longpass filters with cut-on wavelengths of 600 nm and 510 nm, respectively. Linares Rosales et \textit{al.} \cite{Linares-2019} showed that in the wavelength range of 510 to 600 nm, there was scintillation light that was not used. Hence, replacing the A10033-63 filter with the the A10033-71 filter allowed for additional improvement in the signal-collection efficiency in CH-4.

\subsection{Contribution to the uncertainty chain}

\begin{figure}
\centering
\begin{tabular}{c}
\includegraphics[trim = 0mm 0mm 0mm 0mm, clip, scale=0.43]{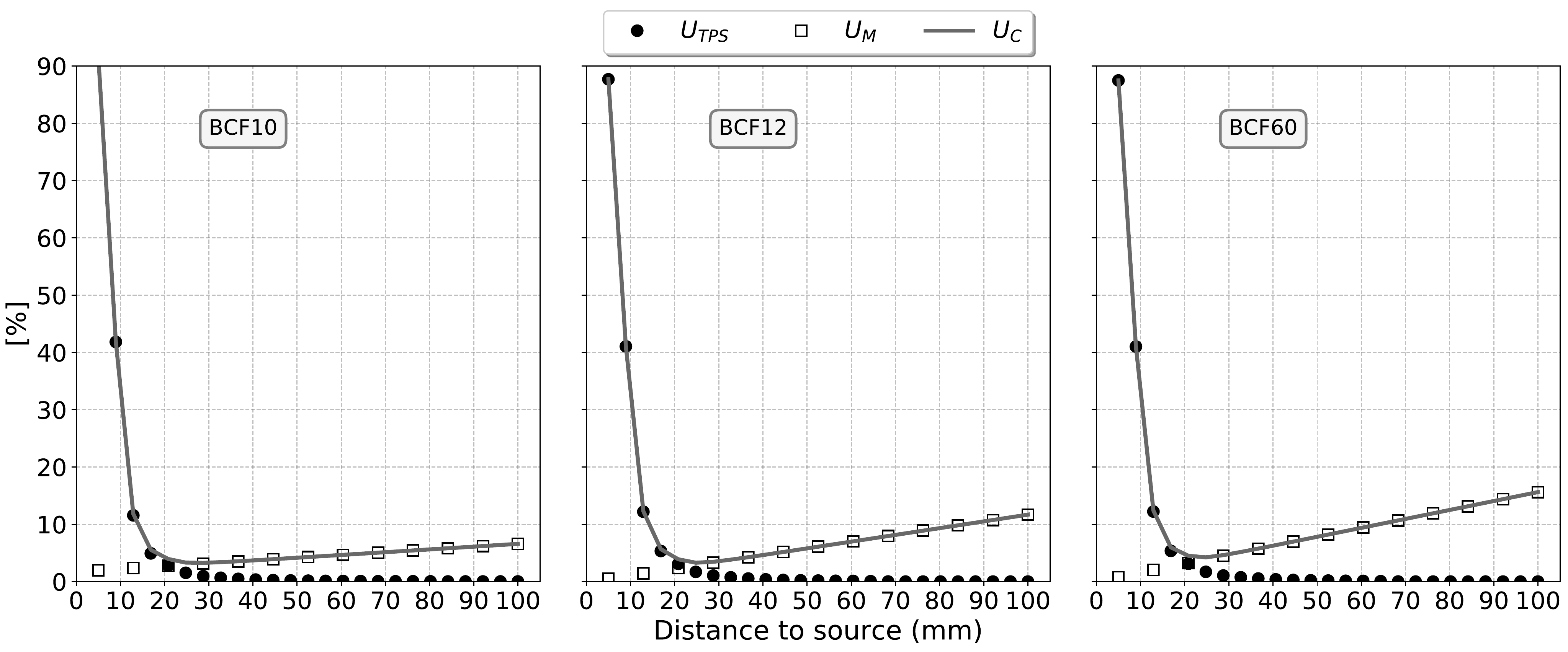} 
\end{tabular}
\caption{\label{mPSD_UTPS_UTG43} Contribution of the measurement and positioning uncertainties to the mPSD response in HDR brachytherapy. The uncertainty values are relative to the dose at the given depth. $U_{c}$ is the combination of the TG-43 dose gradient uncertainty ($U_{TPS}$) and the measurements uncertainty ($U_{M}$).}
\end{figure}

Figure \ref{mPSD_UTPS_UTG43} shows the relative contribution of detector position uncertainty ($U_{TPS}$) and measurement uncertainty ($U_{M}$) as a function of source-to-detector distance for all three scintillators. A similar study was performed by  Andersen \textit{et al.} \cite{Andersen-time-resolved-2009} with an aluminum oxide crystal (Al$_{2}$O$_{3}$:C) attached to a 1-mm optical fiber reporting a maximum $U_{M}$ of about 40 \% at 50 mm from the source.  The maximum $U_{M}$ observed was 17 \% for the BCF-60 at 100 mm from the source. $U_{M}$ values for BCF-10 and BCF-12 were always under 13\% at all the distances tested. Even if the light produced by the scintillators in the mPSD was subject to multiple optical filtration, the measurement uncertainty remained low at the longest distances, especially in comparison with the results reported by Andersen et \textit{al.} \cite{Andersen-time-resolved-2009}. 

It is important to underline that the uncertainty in the expected dose $U_{TPS}$ was solely accounted for by the positioning uncertainty. AAPM Task Group 138 and GEC-ESTRO \cite{TG-138-GEC-ESTRO-2011} reported that the expanded relative propagated uncertainty (k = 2 or 95\% confidence level) for dose at 1 cm of high-energy brachytherapy sources along their transverse plane was 6.8\%. This uncertainty would compound with $U_{C}$ to complete the error chain. 

Figure \ref{mPSD_UTPS_UTG43} constituted a metric in this work to define the most appropriate distance for mPSD calibration. Table \ref{Uncertainty_results} shows the sweet-spot values associated with each scintillator in the mPSD. These distances represent the best compromise between mispositioning and measurement uncertainty for the mPSD system under evaluation. Of course, $U_{C}$ is specific to the detector used—in this case, to each sensor of the multipoint dosimeter. Such analysis should be performed as a standard of practice when reporting the performance of an in vivo dosimeter owing to the strong distance dependence displayed in brachytherapy.

\begin{table}
\caption{\label{Uncertainty_results} Recommended distance to source for mPSD calibration for HDR brachytherapy. }
\centering
\vspace{0.2cm}
	\begin{tabular}{ccc}
	\hline
	\hline
	 & \multicolumn{2}{c}{\textbf{\textit{Sweet-spot}}} \\
	 \textbf{Scintillator} & \textbf{Distance (mm)} & \textbf{$U_c$ (\%)}\\
	\hline
    BCF10           & 17.2      & 3.8 \\
    BCF12           & 17.4      & 3.6 \\
    BCF60           & 17.5      & 4.3 \\
	\hline
	\end{tabular}
\end{table}

\subsection{Angular dependence}
\label{Result_Ang_Dependence}

\begin{figure}
\centering
\begin{tabular}{c}
\includegraphics[trim = 3mm 1mm 1mm 1mm, clip, scale=0.415]{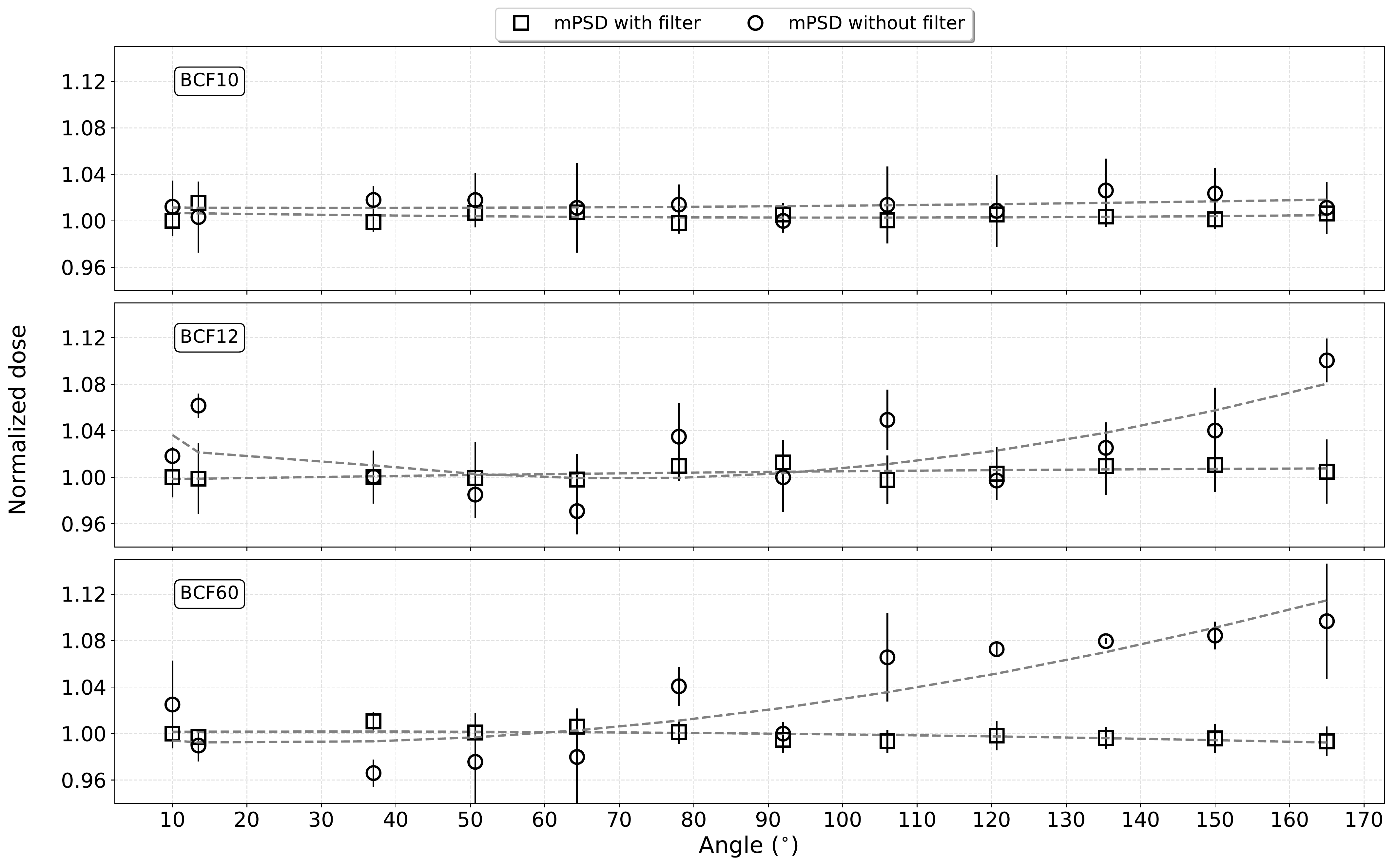}
\end{tabular}
\caption{\label{mPSD_angular_dep} The mPSD’s response as a function of angle to the HDR brachytherapy source. Dose values are normalized to 90$^{\circ}$. Bars represent standard deviations. Dotted lines are the trends in the mPSD’s response.}
\end{figure}

Figure \ref{mPSD_angular_dep} depicts the angular dependence of the mPSD system with and without the use of a bandpass filter coupled to the BCF-60 scintillator. The dotted lines represent the trendlines of each detector’s response. The dose values in Figure \ref{mPSD_angular_dep} are normalized to each scintillator's response at 90$^{\circ}$. Previous studies have analyzed the angular dependence of some plastic scintillators \cite{Archambault-2007, Wang-MC-2010, Wang-MC-2011, Gagnon-2012}, but to our knowledge, none have examined a multipoint detector configuration. A study by Archambault et al. \cite{Archambault-2007} on a single-point plastic scintillator dosimeter composed of BCF-12 and irradiated using an external beam found no angular dependence in response, with a maximum deviation of 0.6\%. They highlighted the importance of employing a stem-effect removal technique to avoid larger deviations caused by angular dependence.  Wang et al. \cite{Wang-MC-2010, Wang-MC-2011} also found angular independence for a BCF-12 detector, with responses varying by about 2\%. Furthermore, the angular independence of a BCF-60 detector has been previously established \cite{Gagnon-2012}. A study by Lambert et al. \cite{Lambert-2006} recommended the use of plastic scintillator dosimeters with diameter-to-length ratios below 5:1 for brachytherapy purposes; this would ensure detector response variation within 1.5\% as a function of angle to the source. The mPSD under evaluation in the present study was composed of 3 mm of BCF-10, 6 mm of BCF-12, and 7 mm of BCF-60. In this context, only the diameter-to-length ratio of BCF-10 would fall into the range recommended by Lambert et \textit{al.} \cite{Lambert-2006}.

Figure \ref{mPSD_angular_dep} shows that as angles went beyond 90$^{\circ}$, a clear angular dependence emerged in the BCF-12 and BCF-60 curves, up to almost +10\% when no filter was used. We hypothesized that this effect was due to cross-excitation of the sensors. We tested this hypothesis by using a 400 to 600-nm bandpass filter coupled to the BCF-60 sensor, which would be the one producing the least amount of direct scintillation light at large angle (farthest from the source) and thus the most susceptible to excitation by the other two scintillators. After this simple addition to the system, all of the scintillator responses were essentially flat at all angles.

\subsection{Source position tracking}

\subsubsection{Absorbed dose measurements}

\begin{figure}
\centering
\begin{tabular}{c}
\includegraphics[trim = 1mm 10mm 10mm 8mm, clip, scale=0.55]{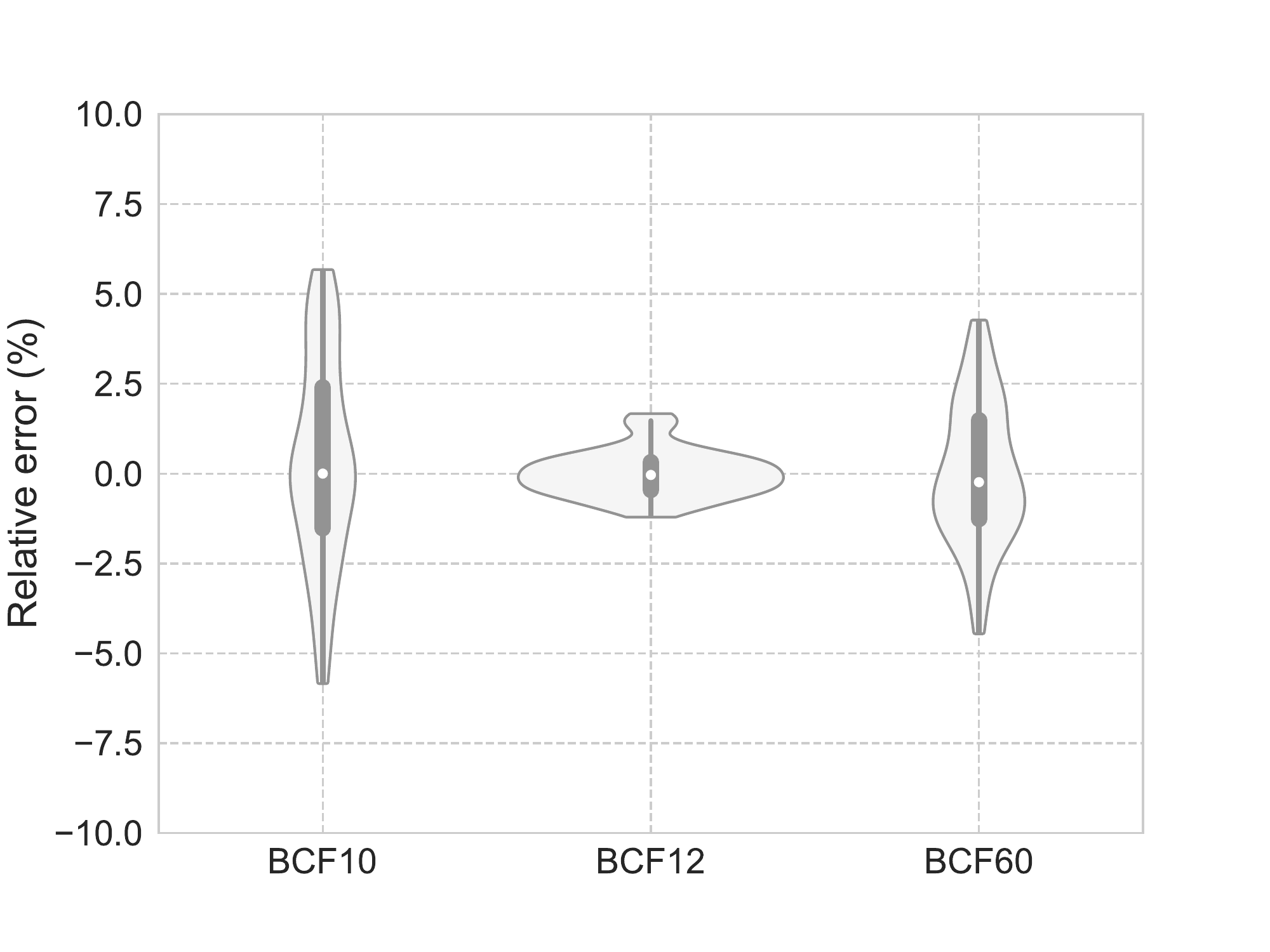} 
\end{tabular}
\caption{\label{source_position_dose} Density distribution of the relative errors between the scintillators’ measured doses and TG-43 during source-tracking analysis for HDR brachytherapy. Boxes represent interquartile ranges (quartile 1 and quartile 3), and the white dots inside the boxes represent the median values. The inner vertical lines extend from each quartile to the minimum or maximum.}
\end{figure}

The violin plots \cite{Hintze-Violin-Plot-1998} in Figure \ref{source_position_dose} show the density distributions of the relative differences between each scintillator’s measured dose and the dose calculated by TG-43 during irradiations with Plans 1 and 2. The inner boxes represent the interquartile ranges, and the white dots inside the boxes indicate the median values. The scintillators' measured doses did not deviate by more than 5\% from the TG-43 predicted dose. The median values were close to 0, and the highest densities were also close to 0 (values close to 0 and a small dispersion around it represent better agreement with the reference).  BCF-12, the middle sensor, had the least dispersion, with an interquartile range within 1\% of the deviations. During measurements, the distance from the BCF-12 detector to the source remained relatively constant owing to its central position inside the mPSD, whereas the BCF-10 and BCF-60 detectors were subject to more extreme distance variations and accordingly exhibited greater dispersion in the difference between measured and expected (TG-43) doses. Nonetheless, for each sensor, 75\% of the sample was found to be within a deviation range of 2.5\%. The points with differences greater than 2.5\% corresponded to positions with source-to-detector distances of more than 6.5 cm.

The high-dose-gradient field imposed by the $^{192}Ir$ source at short distances may induce a substantial uncertainty in the dose determination, on the order of 20\% per millimeter at 10 cm from the source. That effect was not observed in this study. We explain this result thus: the actual position of each sensor in the mPSD was calibrated initially. Also, the expected dose values used in this study were calculated by considering each scintillator as a volume, not as a single point in space. Such an approach accounts for the fact that the dose gradient is not constant and varies as a function of distance to the source, including inside the scintillator volumes themselves.

\subsubsection{Source-position determination}

\begin{figure}
\centering
\begin{tabular}{c}
\includegraphics[trim = 0mm 0mm 0mm 0mm, clip, scale=0.425]{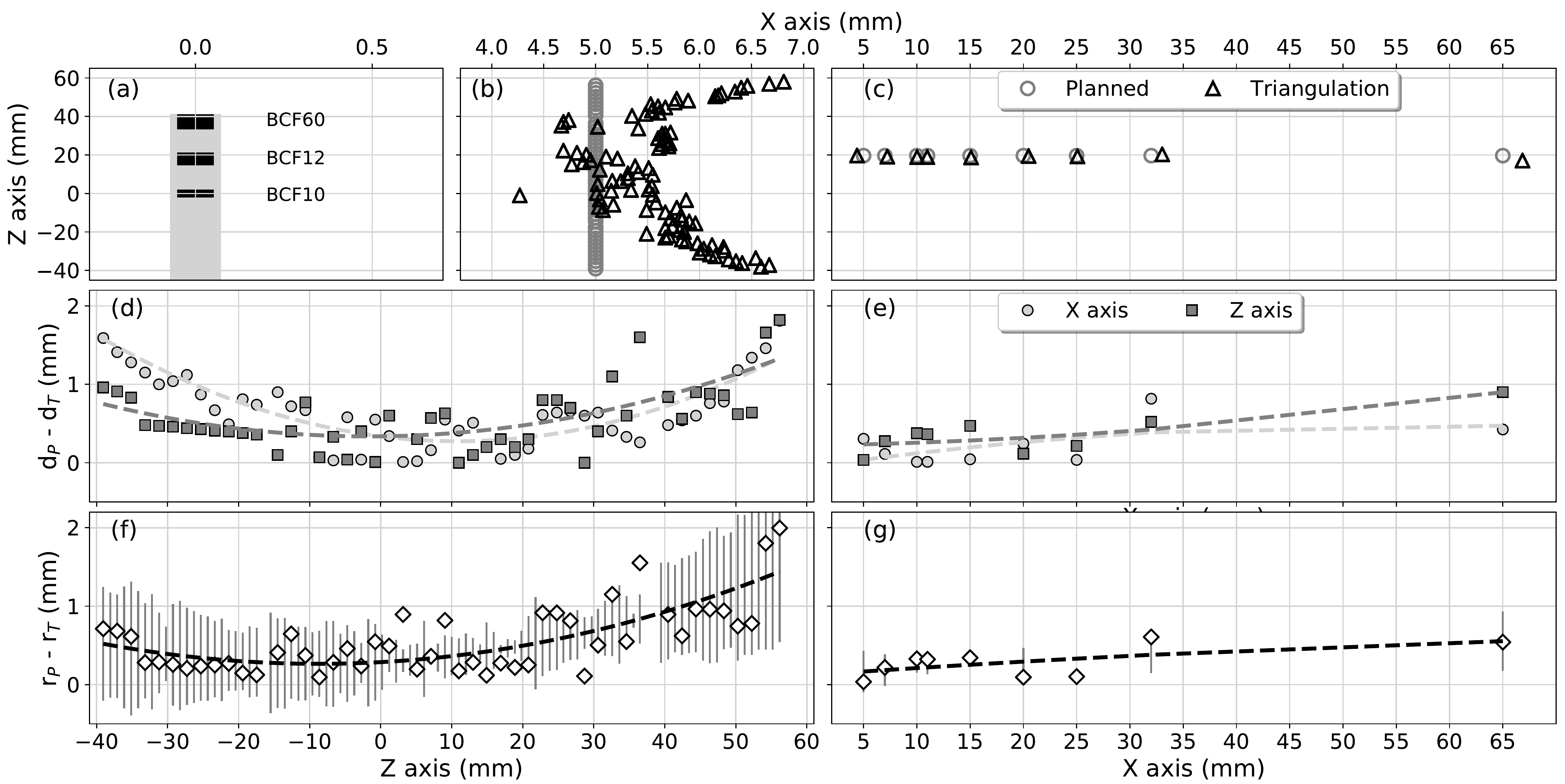} 
\end{tabular}
\caption{\label{source_position_triang} Source-position determination. (a), (b), and (c) illustrate the plans: (a) Schematic of the mPSD coordinate system used for source-position triangulation; (b) corresponds to Plan 1; (c) corresponds to Plan 2. (d) and (e) show differences (in mm) between the distance from the mPSD to the planned source position and the distance to the triangulated one: (d) for Plan 1, (e) for Plan 2. (f) and (g) show the absolute difference between the radius to the planned position and the radius to the triangulated one for Plan 1 (f) and Plan 2 (g). Dotted lines represent the trendlines of the deviations between the calculated positions and the planned ones. In (f) and (g), the vertical gray lines represent the standard deviation of each triangulated position. }
\end{figure}

Figure \ref{source_position_triang} presents the determination of the source position by triangulation \cite{Nakano-2003-HDR-track}. Figures \ref{source_position_triang}a, \ref{source_position_triang}b, and \ref{source_position_triang}c illustrate the configuration of the mPSD and the source positions in the z- and x-directions, respectively, during irradiations. Figure \ref{source_position_triang}a shows the mPSD coordinate system used for source-position triangulation. Figures \ref{source_position_triang}b and \ref{source_position_triang}c illustrate the delivery of Plans 1 and 2, respectively. Figures \ref{source_position_triang}d and \ref{source_position_triang}e summarize the absolute deviation (in mm) on the x and z axes, between the calculated position of the mPSD and the planned one, for a source moving along the z axis (Plan 1) or along the x axis (Plan 2). Figures \ref{source_position_triang}f and \ref{source_position_triang}g show the absolute difference between the radial distance to the planned source position and the triangulated one, along the same two axes. The gray vertical lines in Figures \ref{source_position_triang}f and \ref{source_position_triang}g represent the standard deviation of each triangulated position. The dotted lines in Figures \ref{source_position_triang}d, \ref{source_position_triang}e, \ref{source_position_triang}f, and \ref{source_position_triang}g represent the trendlines of the calculated deviations. 

The source-position tracking showed that, as the source moved away from each scintillator volume, the measurement uncertainty started to affect the mPSD’s capacity to report a precise distance to the source. Thus, the source triangulation process became less effective. The radial distance to the source was defined as the distance from each scintillator’s effective center to the source dwell position. We observed differences above 1 mm in the radial distance prediction at distances past 62 mm for BCF-10, 60 mm for BCF-12, and 55 mm for BCF-60. In addition, BCF-60 had greater $U_M$ values at long distances from the source. Results from Plan 1 (Figures \ref{source_position_triang}b and \ref{source_position_triang}d) demonstrated that when the source dwelled at the extremities of the mPSD, deviations in source reporting could reach a maximum of 1.8 $\pm$ 1.6 mm in the x or z axes. The trendlines shown in Figure \ref{source_position_triang}d help to visualize this behaviour. However, as depicted in Figures \ref{source_position_triang}e and \ref{source_position_triang}g, the maximal observed deviation from the planned position for the delivered Plan 2 was never greater than 0.92 $\pm$ 0.5 mm.

Therriault-Proulx et al. \cite{Therriault-mPSD-Brachy-2013} used a multipoint configuration with a single optical fiber as an in vivo verification tool for HDR brachytherapy. Based on a determination of each independent scintillator’s offset, they used a weighted approach to report the overall offset between the expected and calculated positions of the $^{192}$Ir source. Although this weighted offset improved their source-position detection, offsets greater than 2.5 mm were reported, limiting their HDR brachytherapy measurements to a range within 3 cm of the source. In contrast, the current study demonstrated that an optimized system can extract source positions with maximal deviations of no more than 1.8 mm for a range up to 10 cm from the source.

\subsection{Planned vs. measured dwell time}

\begin{figure}
\centering
\begin{tabular}{c}
\includegraphics[trim = 0mm 0mm 0mm 0mm, clip, scale=0.6]{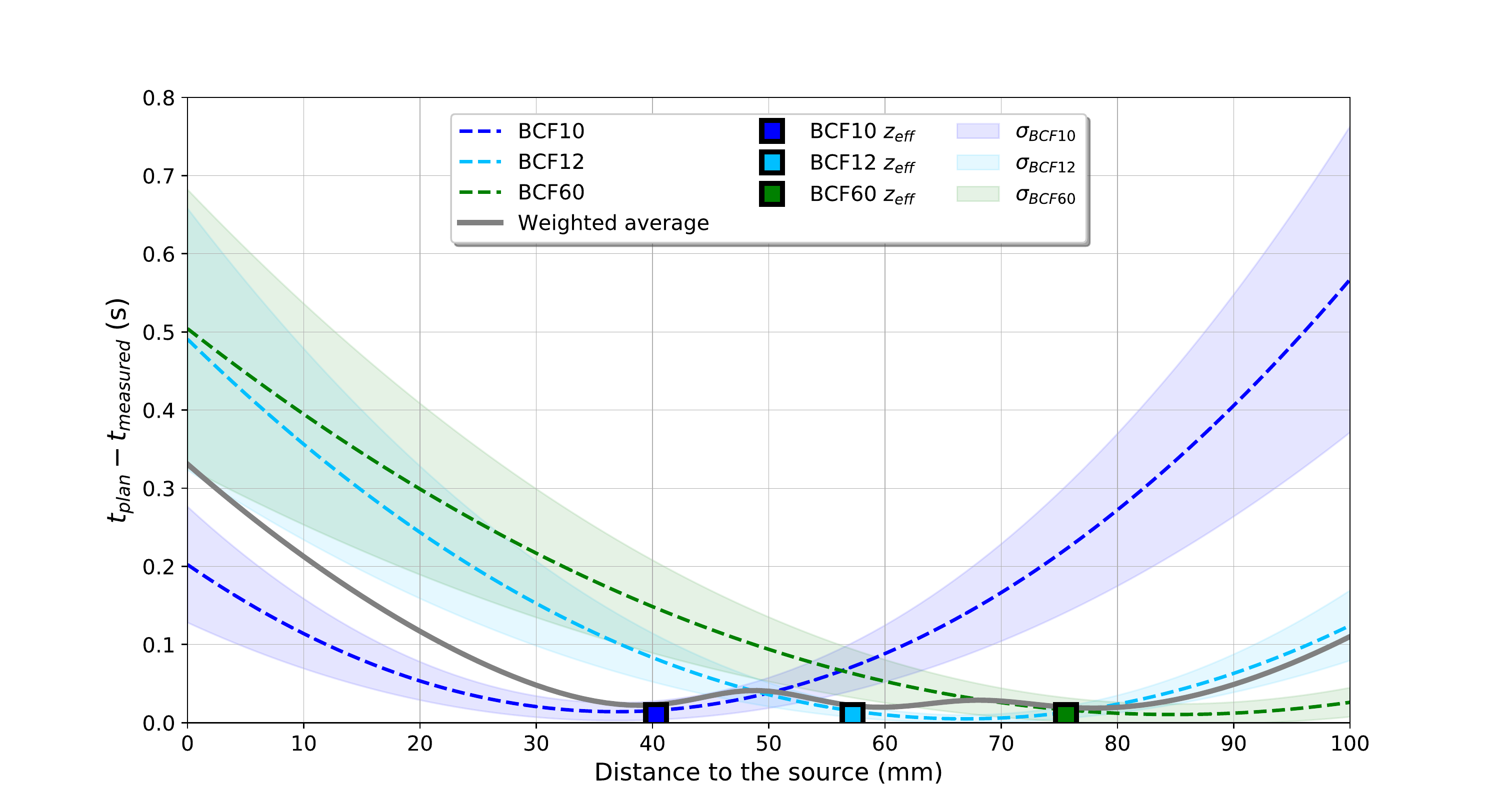}
\end{tabular}
\caption{\label{mPSD_time_res} Deviation of mPSD measured dwell times from planned dwell times as a function of distance to the source. Dotted lines represent the average of the scintillators’ deviations. The continuous line represents the sensors’ weighted average deviation. The shaded region represents the standard deviation. The squares along the bottom correspond to each scintillator’s effective position.}
\end{figure}

Figure \ref{mPSD_time_res} shows the deviations between the measured dwell times for our mPSD system and the planned ones as a function of distance to the source. The dotted lines represent the average measurement differences for each individual scintillator, whose positions are represented by squares. The shaded region around the average line shows the standard deviation extracted from all the dwell times measured. Taking as reference a range from $-$10 to $+$10 mm around each scintillator’s effective position (radially), the average dwell time measured by BCF-10 was within 0.03 $\pm$ 0.04 s of the planned dwell time, while the BCF-12 average response was 0.04 $\pm$ 0.04 s, and that of BCF60 was 0.03 $\pm$ 0.05 s.

When the source was moved to larger distances from a given scintillator’s effective position, the deviations between the measured and planned dwell times increased accordingly. Evidently, at short distances from the source, the high gradient field characteristic of the $^{192}$Ir source allowed to us obtain a sharp pulse of signal and, as a consequence, proper differentiation of the signal from one position to the subsequent one. At long distances, as shown by  Andersen et \textit{al.} \cite{Andersen-time-resolved-2009}, detectable source displacement is more difficult to obtain for small dwell times owing to the increased measurement uncertainty.

The capability of the mPSD system in measuring the source dwell time was evaluated for a range up to 10 cm from the source. The maximum deviation observed at all distances using the BCF-60 detector was 0.56 $\pm$ 0.25 s. The beauty of our mPSD system, however, is that one or more additional sensors could be placed closer to the source to provide an alternate measurement. In our case, a weighted average over all three scintillators was performed (continuous line in Figure \ref{mPSD_time_res}), resulting in a maximum deviation of 0.33 $\pm$ 0.37 s. This level of accuracy is sufficient for clinical validation of individual dwell times for most configurations.

\section{Conclusions}

In this study, we presented the performance of an mPSD system for HDR brachytherapy and studied the uncertainty chain by extracting the relative contributions of measurement and positioning uncertainties as function of distance from the source. We used this analysis as a metric to find the conditions that ensure the best compromise between positioning and measurement uncertainties for mPSD calibration. The mPSD angular response was flat within 2\%, provided that cross re-excitation of the scintillators was prevented by a bandpass filter. The triangulation approach was applied to track the source position in space. As long as the mPSD-to-source distance was within 6 cm, the source position could be extracted to within 1 mm of the expected position, increasing to 1.8 mm at 10 cm. In all of the explored conditions, dose differences relative to TG-43 expected doses were within 5\%. At distances up to 6.5 cm the dose deviation distribution for each sensor was within 2.5\% of the TG-43 expected dose. The mPSD’s capacity for measuring the source dwell time was evaluated for a range of source-to-detector distances of up to 10 cm, with a maximum single-scintillator deviation of 0.56 $\pm$ 0.25 s. Dwell time measurements exhibited the largest deviations for small dwell time values (1 s) and longer distances from the scintillator’s effective position. However, the maximum average weighted deviation of the measured dwell times over all three scintillators was 0.33 $\pm$ 0.37 s, and the weighted average was 0.07 $\pm$ 0.05 s at all distances covered in this study. Thus, this study demonstrated that the described system can be used as an in vivo dosimeter for real-time source tracking, individual dwell time measurements, and dose reporting.

\begin{acknowledgments}

The present work was supported by the National Sciences and Engineering Research Council of Canada (NSERC) via the NSERC-Elekta Industrial Research Chair grants Nos. 484144-15 and RGPIN-2019-05038, and by a Canadian Foundation for Innovation (CFI) JR Evans Leader Funds grant \# 35633. Haydee Maria Linares Rosales further acknowledges support from Fonds de Recherche du Quebec - Nature et Technologies (FRQ-NT) and by the CREATE Medical Physics Research Training Network grant of the Natural Sciences and Engineering Research Council of Canada (Grant \# 432290). We also thank Amy Ninetto from Scientific Publications Services in the Research Medical Library at The University of Texas MD Anderson Cancer Center for editing our manuscript.

\end{acknowledgments}


\begin{thebibliography}{10}

\bibitem{Kertzscher-2011}
G.~Kertzscher, C.~E. Andersen, F.-A. Siebert, S.~K. Nielsen, J.~C. Lindegaard,
  and K.~Tanderup, ``Identifying afterloading {PDR} and {HDR} brachytherapy
  errors using real-time fiber-coupled {Al}2o3:{C} dosimetry and a novel
  statistical error decision criterion,'' {\em Radiotherapy and Oncology},
  vol.~100, pp.~456--462, Sept. 2011.
  
 \bibitem{ICRP-97-Brachy}
J.~Valentin, ``Icrp publication 97: Prevention of high-dose-rate brachytherapy
  accidents,'' {\em Annals of the ICRP}, vol.~35, pp.~1--51, 02 2005.
  
\bibitem{DAS-TLD-invivo-2007}
R.~Das, W.~Toye, T.~Kron, S.~Williams, and G.~Duchesne, ``Thermoluminescence
  dosimetry for in-vivo verification of high dose rate brachytherapy for
  prostate cancer,'' {\em Australasian Physics \& Engineering Sciences in
  Medicine}, vol.~30, p.~178, Sept. 2007.
  
\bibitem{Toye-TLD-invivo-2009}
W.~Toye, R.~Das, T.~Kron, R.~Franich, P.~Johnston, and G.~Duchesne, ``An in
  vivo investigative protocol for {HDR} prostate brachytherapy using urethral
  and rectal thermoluminescence dosimetry,'' {\em Radiotherapy and Oncology},
  vol.~91, pp.~243--248, May 2009.
  
\bibitem{Anagnostopoulos-TLD-invivo-2003}
G.~Anagnostopoulos, D.~Baltas, A.~Geretschlaeger, T.~Martin, P.~Papagiannis,
  N.~Tselis, and N.~Zamboglou, ``In vivo thermoluminescence dosimetry dose
  verification of transperineal 192ir high-dose-rate brachytherapy using
  {CT}-based planning for the treatment of prostate cancer,'' {\em
  International Journal of Radiation Oncology*Biology*Physics}, vol.~57,
  pp.~1183--1191, Nov. 2003.
  
\bibitem{Tanderup-invivo-Brachy-2013}
K.~Tanderup, S.~Beddar, C.~E. Andersen, G.~Kertzscher, and J.~E. Cygler, ``In
  vivo dosimetry in brachytherapy,'' {\em Medical Physics}, vol.~40, no.~7,
  p.~070902, 2013.
  
\bibitem{Therriault-Temp-method-2015}
F.~Therriault-Proulx, L.~Wootton, and S.~Beddar, ``A method to correct for
  temperature dependence and measure simultaneously dose and temperature using
  a plastic scintillation detector,'' {\em Physics in Medicine and Biology},
  vol.~60, no.~20, p.~7927, 2015.

\bibitem{Boivin-2016}
J.~Boivin, S.~Beddar, C.~Bonde, D.~Schmidt, W.~Culberson, M.~Guillemette, and
  L.~Beaulieu, ``A systematic characterization of the low-energy photon
  response of plastic scintillation detectors,'' {\em Physics in Medicine and
  Biology}, vol.~61, no.~15, p.~5569, 2016.
  
\bibitem{Lambert-Cerenkov-2008}
J.~Lambert, Y.~Yin, D.~R. McKenzie, S.~Law, and N.~Suchowerska, ``Cerenkov-free
  scintillation dosimetry in external beam radiotherapy with an air core light
  guide,'' {\em Physics in Medicine and Biology}, vol.~53, no.~11, p.~3071,
  2008.

\bibitem{Beddar-Cerenkov-1992}
A.~S. Beddar, T.~R. Mackie, and F.~H. Attix, ``Cerenkov light generated in
  optical fibres and other light pipes irradiated by electron beams,'' {\em
  Physics in Medicine and Biology}, vol.~37, no.~4, pp.~925--935, 1992.
  
\bibitem{Archambault-2006}
L.~Archambault, A.~S. Beddar, L.~Gingras, R.~Roy, and L.~Beaulieu,
  ``Measurement accuracy and cerenkov removal for high performance, high
  spatial resolution scintillation dosimetry,'' {\em Medical Physics}, vol.~33,
  no.~1, pp.~128--135, 2006.

\bibitem{Wootton-Temperature-2013}
L.~Wootton and S.~Beddar, ``Temperature dependence of bcf plastic scintillation
  detectors,'' {\em Physics in Medicine and Biology}, vol.~58, no.~9, p.~2955,
  2013.
  
\bibitem{Guillot-toward-2010}
M.~Guillot, L.~Gingras, L.~Archambault, S.~Beddar, and L.~Beaulieu, ``Toward 3d
  dosimetry of intensity modulated radiation therapy treatments with plastic
  scintillation detectors,'' {\em Journal of Physics: Conference Series},
  vol.~250, no.~1, p.~012006, 2010.

\bibitem{Beddar-water-equivalent-1992-1}
A.~S. Beddar, T.~R. Mackie, and F.~H. Attix, ``Water-equivalent plastic
  scintillation detectors for high-energy beam dosimetry: {I}. {Physical}
  characteristics and theoretical considerations,'' {\em Physics in Medicine
  and Biology}, vol.~37, no.~10, pp.~1883--1900, 1992.
  
\bibitem{Beddar-water-equivalent-1992-2}
A.~S. Beddar, T.~R. Mackie, and F.~H. Attix, ``Water-equivalent plastic
  scintillation detectors for high-energy beam dosimetry: {II}. {Properties}
  and measurements,'' {\em Physics in Medicine and Biology}, vol.~37, no.~10,
  p.~1901, 1992.
  
\bibitem{Beaulieu-Scint-Status-2013}
L.~Beaulieu, M.~Goulet, L.~Archambault, and S.~Beddar, ``Current status of
  scintillation dosimetry for megavoltage beams,'' {\em Journal of Physics:
  Conference Series}, vol.~444, no.~1, p.~012013, 2013.
  
\bibitem{Beaulieu-Review-2016}
L.~Beaulieu and S.~Beddar, ``Review of plastic and liquid scintillation
  dosimetry for photon, electron, and proton therapy,'' {\em Physics in
  Medicine and Biology}, vol.~61, no.~20, p.~R305, 2016.
  
\bibitem{Linares-2019}
H.~M. Linares~Rosales, P.~Duguay-Drouin, L.~Archambault, S.~Beddar, and
  L.~Beaulieu, ``Optimization of a multipoint plastic scintillator dosimeter
  for high dose rate brachytherapy,'' {\em Medical Physics}, vol.~46, no.~5,
  pp.~2412--2421, 2019.
  
\bibitem{Beddar-temp-2012}
S.~Beddar, ``On possible temperature dependence of plastic scintillator
  response,'' {\em Medical Physics}, vol.~39, no.~10, pp.~6522--6522, 2012.
  
\bibitem{Boer-optical-1993}
S.~F. de~Boer, A.~S. Beddar, and J.~A. Rawlinson, ``Optical filtering and
  spectral measurements of radiation-induced light in plastic scintillation
  dosimetry,'' {\em Physics in Medicine and Biology}, vol.~38, no.~7, p.~945,
  1993.
  
\bibitem{Fontbonne-2002}
J.~M. Fontbonne, G.~Iltis, G.~Ban, A.~Battala, J.~C. Vernhes, J.~Tillier,
  N.~Bellaize, C.~L. Brun, B.~Tamain, K.~Mercier, and J.~C. Motin,
  ``Scintillating fiber dosimeter for radiation therapy accelerator,'' {\em
  IEEE Transactions on Nuclear Science}, vol.~49, no.~5, pp.~2223--2227, 2002.
  
\bibitem{Clift-temporal-2002}
M.~A. Clift, P.~N. Johnston, and D.~V. Webb, ``A temporal method of avoiding
  the cerenkov radiation generated in organic scintillator dosimeters by pulsed
  mega-voltage electron and photon beams,'' {\em Physics in Medicine amd
  Biology}, vol.~47, no.~8, p.~1421, 2002.
  
\bibitem{Archambault-MathForm-2012}
L.~Archambault, F.~Therriault-Proulx, S.~Beddar, and L.~Beaulieu, ``A
  mathematical formalism for hyperspectral, multipoint plastic scintillation
  detectors,'' {\em Physics in Medicine and Biology}, vol.~57, no.~21, p.~7133,
  2012.
  
\bibitem{Therriault-mPSD-2012}
F.~Therriault-Proulx, L.~Archambault, L.~Beaulieu, and S.~Beddar, ``Development
  of a novel multi-point plastic scintillation detector with a single optical
  transmission line for radiation dose measurement,'' {\em Physics in Medicine
  and Biology}, vol.~57, no.~21, p.~7147, 2012.
 
\bibitem{Therriault-mPSD-Brachy-2013}
F.~Therriault-Proulx, S.~Beddar, and L.~Beaulieu, ``On the use of a
  single-fiber multipoint plastic scintillation detector for {$^{192}$Ir}
  high-dose-rate brachytherapy,'' {\em Medical Physics}, vol.~40, no.~6,
  p.~062101, 2013.
  
\bibitem{Johansen-2018}
J.~G. Johansen, S.~Rylander, S.~Buus, L.~Bentzen, S.~B. Hokland, C.~S.
  Søndergaard, A.~K.~M. With, G.~Kertzscher, and K.~Tanderup, ``Time-resolved
  in vivo dosimetry for source tracking in brachytherapy,'' {\em
  Brachytherapy}, vol.~17, pp.~122--132, Feb. 2018.

\bibitem{Smith-2016}
R.~L. Smith, A.~Haworth, V.~Panettieri, J.~L. Millar, and R.~D. Franich, ``A
  method for verification of treatment delivery in hdr prostate brachytherapy
  using a flat panel detector for both imaging and source tracking,'' {\em
  Medical Physics}, vol.~43, no.~5, pp.~2435--2442.
  
\bibitem{Guiral-2016}
P.~Guiral, J.~Ribouton, P.~Jalade, R.~Wang, J.-M. Galvan, G.-N. Lu, P.~Pittet,
  A.~Rivoire, and L.~Gindraux, ``Design and testing of a phantom and
  instrumented gynecological applicator based on gan dosimeter for use in high
  dose rate brachytherapy quality assurance,'' {\em Medical Physics}, vol.~43,
  no.~9, pp.~5240--5251.

\bibitem{Nakano-2005}
T.~Nakano, N.~Suchowerska, D.~R. McKenzie, and M.~M. Bilek, ``Real-time
  verification of hdr brachytherapy source location: implementation of detector
  redundancy,'' {\em Physics in Medicine and Biology}, vol.~50, no.~2, p.~319,
  2005.
  
\bibitem{Fonseca-2017}
G.~P. Fonseca, M.~Podesta, M.~Bellezzo, M.~R.~V. den Bosch, L.~Lutgens,
  B.~G.~L. Vanneste, R.~Voncken, E.~J.~V. Limbergen, B.~Reniers, and
  F.~Verhaegen, ``Online pretreatment verification of high-dose rate
  brachytherapy using an imaging panel,'' {\em Physics in Medicine and
  Biology}, vol.~62, no.~13, p.~5440, 2017.

\bibitem{Kertzscher-2016-Ruby}
G.~Kertzscher and S.~Beddar, ``Ruby-based inorganic scintillation detectors
  for192ir brachytherapy,'' {\em Physics in Medicine and Biology}, vol.~61,
  pp.~7744--7764, oct 2016.
  
\bibitem{Therriault-2011}
F.~Therriault-Proulx, S.~Beddar, T.~M. Briere, L.~Archambault, and L.~Beaulieu,
  ``Technical note: Removing the stem effect when performing {Ir-192} {HDR}
  brachytherapy in vivo dosimetry using plastic scintillation detectors: A
  relevant and necessary step,'' {\em Medical Physics}, vol.~38, no.~4,
  pp.~2176--2179, 2011.
  
 \bibitem{Ayotte-Surface-2006}
G.~Ayotte, L.~Archambault, L.~Gingras, F.~Lacroix, A.~S. Beddar, and
  L.~Beaulieu, ``Surface preparation and coupling in plastic scintillator
  dosimetry,'' {\em Medical Physics}, vol.~33, no.~9, pp.~3519--3525, 2006.
  
\bibitem{PMT-Hamamatsu}
{Hamamatsu Photonics}, ``Hamamatsu {PMT} {H10722} {Series},'' October 2016.
\newblock
  \url{https://www.hamamatsu.com/resources/pdf/etd/H10722_TPMO1063E.pdf}.
  
\bibitem{DAQ-6289}
{National Instruments Corporation}, ``National {Instruments} {Multifunction}
  {Data} {Acquisition} {Device},'' October 2016.
\newblock \url{http://sine.ni.com/nips/cds/view/p/lang/en/nid/209154}.

\bibitem{TG-43-Update}
M.~J. Rivard, B.~M. Coursey, L.~A. DeWerd, W.~F. Hanson, H.~M. Saiful, G.~S.
  Ibbott, M.~G. Mitch, R.~Nath, and J.~F. Williamson, ``Update of aapm task
  group no. 43 report: A revised aapm protocol for brachytherapy dose
  calculations,'' {\em Medical Physics}, vol.~31, no.~3, pp.~633--674, 2004.
  
\bibitem{Andersen-time-resolved-2009}
C.~E. Andersen, S.~K. Nielsen, J.~C. Lindegaard, and K.~Tanderup,
  ``Time-resolved in vivo luminescence dosimetry for online error detection in
  pulsed dose-rate brachytherapy,'' {\em Medical Physics}, vol.~36, no.~11,
  pp.~5033--5043, 2009.
  
\bibitem{Perez-2012-HEBD}
J.~Perez-Calatayud, F.~Ballester, R.~K. Das, L.~A. DeWerd, G.~S. Ibbott, A.~S.
  Meigooni, Z.~Ouhib, M.~J. Rivard, R.~S. Sloboda, and J.~F. Williamson, ``Dose
  calculation for photon-emitting brachytherapy sources with average energy
  higher than 50 kev: Report of the aapm and estro,'' {\em Medical Physics},
  vol.~39, no.~5, pp.~2904--2929, 2012.
  
\bibitem{TG-138-GEC-ESTRO-2011}
L.~A. DeWerd, G.~S. Ibbott, A.~S. Meigooni, M.~G. Mitch, M.~J. Rivard, K.~E.
  Stump, B.~R. Thomadsen, and J.~L.~M. Venselaar, ``A dosimetric uncertainty
  analysis for photon-emitting brachytherapy sources: Report of aapm task group
  no. 138 and gec-estro,'' {\em Medical Physics}, vol.~38, no.~2, pp.~782--801,
  2011.
  
\bibitem{Archambault-2007}
L.~Archambault, A.~S. Beddar, L.~Gingras, F.~Lacroix, R.~Roy, and L.~Beaulieu,
  ``Water-equivalent dosimeter array for small-field external beam
  radiotherapy,'' {\em Medical Physics}, vol.~34, pp.~1583--1592, May 2007.
  
\bibitem{Wang-MC-2011}
L.~L.~W. Wang and S.~Beddar, ``Study of the response of plastic scintillation
  detectors in small-field 6 {MV} photon beams by {Monte} {Carlo}
  simulations,'' {\em Medical Physics}, vol.~38, pp.~1596--1599, Mar. 2011.

\bibitem{Wang-MC-2010}
L.~L.~W. Wang, D.~Klein, and A.~S. Beddar, ``Monte {Carlo} study of the energy
  and angular dependence of the response of plastic scintillation detectors in
  photon beams,'' {\em Medical Physics}, vol.~37, pp.~5279--5286, Oct. 2010.
  
\bibitem{Gagnon-2012}
J.-C. Gagnon, D.~Thériault, M.~Guillot, L.~Archambault, S.~Beddar, L.~Gingras,
  and L.~Beaulieu, ``Dosimetric performance and array assessment of plastic
  scintillation detectors for stereotactic radiosurgery quality assurance,''
  {\em Medical Physics}, vol.~39, pp.~429--436, Jan. 2012.

\bibitem{Lambert-2006}
J.~Lambert, D.~R. McKenzie, S.~Law, J.~Elsey, and N.~Suchowerska, ``A plastic
  scintillation dosimeter for high dose rate brachytherapy,'' {\em Physics in
  Medicine and Biology}, vol.~51, no.~21, p.~5505, 2006.

\bibitem{Hintze-Violin-Plot-1998}
J.~L. Hintze and R.~D. Nelson, ``Violin plots: A box plot-density trace
  synergism,'' {\em The American Statistician}, vol.~52, no.~2, pp.~181--184,
  1998.
  
\bibitem{Nakano-2003-HDR-track}
T.~Nakano, N.~Suchowerska, M.~M. Bilek, D.~R. McKenzie, N.~Ng, and T.~Kron,
  ``High dose-rate brachytherapy source localization: positional resolution
  using a diamond detector,'' {\em Physics in Medicine and Biology}, vol.~48,
  pp.~2133--2146, jul 2003.


\end{thebibliography}

\newpage

\end{document}